\documentstyle[11pt,aaspp4,psfig]{article}

\begin{document}     
\def\today{\ifcase\month\or
January\or February\or March\or April\or May\or June\or
July\or August\or September\or October\or November\or December\fi
\space\number\day, \number\year}

\newcommand{\squig}{$\sim$}
\newcommand{\squigleq}{\mbox{$^{<}\mskip-10.5mu_\sim$}}
\newcommand{\squiggeq}{\mbox{$^{>}\mskip-10.5mu_\sim$}}
\newcommand{\squiggeqmm}{\mbox{$^{>}\mskip-10.5mu_\sim$}}
\newcommand{\decsec}[2]{$#1\mbox{$''\mskip-7.6mu.\,$}#2$}
\newcommand{\decsecmm}[2]{#1\mbox{$''\mskip-7.6mu.\,$}#2}
\newcommand{\decdeg}[2]{$#1\mbox{$^\circ\mskip-6.6mu.\,$}#2$}
\newcommand{\decdegmm}[2]{#1\mbox{$^\circ\mskip-6.6mu.\,$}#2}
\newcommand{\decsectim}[2]{$#1\mbox{$^{\rm s}\mskip-6.3mu.\,$}#2$}
\newcommand{\decmin}[2]{$#1\mbox{$'\mskip-5.6mu.$}#2$}
\newcommand{\asecbyasec}[2]{#1$''\times$#2$''$}
\newcommand{\aminbyamin}[2]{#1$'\times$#2$'$}

\title{The Meinunger ``Nicht Rote" Objects}
\author{Bruce Margon and Eric W. Deutsch}
\affil{Department of Astronomy, 
       University of Washington, Box 351580,
       Seattle, WA 98195-1580\\
       margon@astro.washington.edu; deutsch@astro.washington.edu}

\begin{center}
Accepted for publication in \\
{\it Publications of the Astronomical Society of the Pacific} \\
Volume 111, January 1999
\end{center}

\begin{abstract}

Four high-latitude slow variable stars have been noted by Meinunger (1972) as 
``nicht rote" (``not red") objects and thus curious. We have previously reported 
(Margon \& Deutsch 1997) that one of these objects, CC~Boo, is in fact a QSO.
Here we present observations demonstrating that the remaining three are also 
highly variable active galactic nuclei. The most interesting object of the 
four is perhaps S~10765 (= NGP9~F324-0276706), which proves to be a resolved
galaxy at $z=0.063$. Despite the rapid and large reported variability
amplitude ($\Delta m\sim1.6$~mag), the spectrum is that of a perfectly normal
galaxy, with no emission lines or evident nonthermal continuum. We also
present new spectroscopic and photometric observations for AR~CVn, suggested
by Meinunger
to be an RR~Lyrae star despite its very faint magnitude ($<$$B$$>$=19.4). The
object is indeed one of the most distant RR~Lyrae stars known, at a
galactocentric distance of $\sim40$~kpc.

\end{abstract}

\clearpage
\section{INTRODUCTION}

During a photographic survey 
for variable stars within the globular 
cluster M3, Meinunger (1972) also noted the presence of a handful of
miscellaneous variables not obviously connected with the cluster. One of these
stars, S~10762, now known as CC~Boo, proves instead to be a quasi-stellar
object (Margon \& Deutsch 1997) of $z=0.172$. This QSO was also independently 
discovered as the optical counterpart of a ROSAT X-ray source by Lamer et al.
(1997), but they were apparently unaware of its association with the cataloged
variable star. Here we discuss observations of four other objects in this
sample, each of which proves to be interesting.

CC~Boo was one of four variables noted by Meinunger as ``nicht rote" (``not 
red") stars. As high-latitude slow variables are often red
giants, this color curiosity is immediately of interest. Each of the
remaining three
``nicht rote" objects indeed proves to be something other than an LPV, and 
each is also extragalactic. The
final variable from the M3 field discussed here, AR~CVn, proves to be one of
the most distant halo stars known.

The variable star literature of course has previous examples of objects 
which ultimately prove to be extragalactic. The most famous case may be 
BL~Lac (Schmitt 1968). Other well-known examples are discussed, e.g., by Bond
(1971, 1973) and Bond \& Tifft (1974).

Our observations were made with the 3.5m
telescope of the Astrophysical Research Consortium, located on Apache Point,
New Mexico. We employed the Double Imaging Spectrograph (DIS), which can
function as both a camera and spectrograph, for most of the observations.
The spectra have resolution $\lambda/\Delta\lambda\sim400$. 

We have also measured accurate ($\pm1''$) positions for the
objects, based on the STScI {\it Digitized Sky Survey.}\footnote{The Digitized
Sky Surveys were produced at the Space Telescope Science Institute under U. S.
Government grant NAGW-2166.}
These improved positions are important, as 
those tabulated by Meinunger (1972) prove to have errors as large as 
\decmin{1}{5}, and this has probably inhibited previous cross-identifications 
with objects which we now know to be interesting.

\section{THE ``nicht rote" VARIABLES}

\subsection{S~10763 = CD Boo}

S~10763 is tabulated (Meinunger 1972) as exhibiting $19.0 < B < 19.7$. It was
subsequently named CD~Boo (Kukarkin et al. 1975, Kholopov et al. 1985). The
only explicit mention of this object in the literature of which we are aware
is that of
Jura \& Kleinmann (1992), who included the object in a list of candidate halo
red giants, based on the GCVS classification; this interpretation is then 
carried forward into the SIMBAD database. However, CD~Boo is undoubtedly 
identical to the published quasar candidate 1339.1+2804 (Crampton et 
al. 1988, 1989), selected for peculiar continuum color from slitless spectra 
plates, where it is noted as ``either a QSO or blue star," 
but with no measurable
redshift.  The coincidence of these objects has presumably escaped previous 
attention in part due to the poor coordinates of S~10763. Newly measured, 
accurate coordinates are given in Table~1.

We obtained the spectrum of CD~Boo on 1998 January 30; 
the resulting data appear 
in Figure~1. The object proves to be a 
QSO with $z=1.045$, based on detection of emission from C\,III] $\lambda$1909,
Mg\,II $\lambda$2800, and $H\gamma$. Thus oddly both CC~Boo (Margon \& Deutsch
1997) and CD~Boo have similar histories: cataloged as halo red giants, but 
actually highly variable QSOs.

CD~Boo appears to be detected at 20~cm in the FIRST survey (Becker et al.
1995, White et al. 1997) with a flux of 1.21~mJy. Unlike CC~Boo, there is no
evidence for X-ray emission in the ROSAT All Sky Survey Bright Source Catalog
(Voges et al. 1998),
but as this object is appreciably
fainter than CC~Boo, the implied $f_x/f_{opt}$ ratio is still within a normal 
range. For $H_0=75$~km~s$^{-1}$~Mpc$^{-1}$ and $q_0=0$ (assumed throughout), 
$B_{max}=19$ implies $M_B=-23.5$, quite typical for a QSO.

\subsection{S~10764 = B2 1339+28 = 1E 1339.9+2844}

The variable S~10764 is observed by Meinunger to have 
$18.3 < B < 19.8$, a particularly large amplitude. Accurate coordinates are 
given in Table~1, and reveal that this object is actually well-studied at 
multiple other wavelengths, but not previously recognized 
as associated with the variable 
star, presumably at least in part due to poor coordinates in the discovery 
paper. The object is associated with the 
strong double-lobed radio source  B2~1339+28 (= 87GB~133954.7+284413), and on
the basis of positional coincidence and marked UV excess, $(U-B)=-1.0$, was
proposed as a QSO by Carney (1976). It is also the optical counterpart of the
X-ray source 1E~1339.9+2844: a redshift of $z=0.33$ is reported by Harris et
al. (1992), who include it in a sample of X-ray selected QSOs near globular
clusters and thus potentially useful for astrometry. This curiously brings
the problem full-circle, as the original discovery of S~10764 was due to its
proximity to M3. These authors as well as Carney (1976) quote magnitudes
compatible with that seen by Meinunger (1972). Additional X-ray observations
from ROSAT are reported by Hertz et al. (1993), where the source is termed
RX\,J134211+2828.7. The published $B_{max}=18.3$ implies $M_B=-22.0$, again a
typical value.

\subsection{S~10765 = NGP9~F324-0276706}

This object appears never to have been mentioned in the literature subsequent 
to the discovery paper, where photometry ranging over 
$17.4 < B < 19.0$ is reported. 
However accurate coordinates (Table~1) reveal it to be identical to the 
cataloged galaxy NGP9~F324-0276706, which appears in NED\footnote{The 
NASA/IPAC Extragalactic Database (NED) is operated by the Jet Propulsion 
Laboratory, California Institute of Technology, under contract with NASA.} 
as derived from the 
Minnesota digitization of POSS-I described by Odewahn \& Aldering (1995).
The object does not appear in the ROSAT All Sky
Survey Bright Source Catalog 
or FIRST radio survey, although it lies in 
regions covered by both. A 20~cm upper limit of 0.97~mJy/beam may be inferred.
We obtained
the spectrum on 1998 January 30. 
Even with the small spectrograph slit-viewing camera 
it was immediately evident that this object is well-resolved, with a 
relatively high surface brightness for at least several arcseconds. The 
resulting spectrum (Figure~2) is that of a normal galaxy at $z=0.063$.

Further imaging observations were made with the 3.5m telescope and SPIcam,
a CCD camera utilizing a thinned SITe 2048 pixel chip of scale 
\decsec{0}{28}~pixel$^{-1}$. All data were obtained with an $R$ filter,
including 
observations on a night of photometric quality when a
standard star was also observed. The images again clearly show the object is
well-resolved, with an extent of many arcsec evident even on casual
inspection. From the photometric night, we derive 
$R=16.65\pm0.10$
within a circle of radius
\decsec{5}{5}, which encompasses the majority of the light.
The modest precision estimate
due in part to the lack of multiple standards.
Relative photometry from data obtained on 2 other nights, for a total
span of $\sim4$ months in 1998,
are also consistent
with this magnitude, {\it i.e.}, we have not yet seen the object vary.
Meinunger quotes $B$ magnitudes for the object, but any values of $(B-R)$ 
consistent with our observed spectrum would imply that our photometry and 
spectroscopy were obtained at or near maximum light. 

S~10765 exhibits an interesting and uncommon combination of properties.
Meinunger reports variations of $\sim1^m$ or more 
on timescales of weeks. Certainly
such behavior is seen in galaxies with active nuclei such as QSOs and
BL~Lacertae objects, but in these cases one typically expects the spectrum to
either show emission lines at minimum light, or weak or absent absorption at
maximum light due to dilution by a nonthermal continuum, together with an
unresolved, long-lived radio component. However, even if the stringent
observed FIRST upper limit on radio emission applies to the object in its
optical minimum state, the inferred ratio of radio to optical luminosity is
far less than appropriate for BL~Lac's (e.g., Stickel et al. 1991). Tempered
only slightly by the lack of contemporaneous radio and optical observations,
we conclude S~10765 exhibits none of the features common to AGN.
Indeed, our observed $R=16.6$, inferred to be measured near to 
{\it maximum} light,
implies $M_{R,max}=-20.3$, hardly typical for a rapidly variable AGN.

As perhaps the least unusual explanation, we have considered the possibility 
that the published finding chart has marked the incorrect object as the 
variable. However our multiple photometric observations have 
included measurement of all 
objects within $\sim2'$ of the galaxy, and none is variable in our data.
Another possibility is the well-known phenomenon that the measured 
apparent magnitude 
of a slightly extended object with respect to nearby stars can depend on 
seeing (e.g., Bond et al. 1974, Green et al. 1977), although in this case the 
reported amplitude of variation is quite large.
Clearly this object is deserving of further study.

\section{AR CVn: AN EXTREMELY DISTANT HALO RR LYRAE STAR}

We have also performed extensive observations of
another of the Meinunger
sample of variables near M3, S~10760, now known (Kukarkin et al. 1975, Kholopov
et al. 1985)
as AR~CVn.
On the basis of the light curve shape and a derived
period of 0.62722~day, the object was classified by Meinunger 
as an RR~Lyrae variable,
with a comment on its faintness ($18.9<B<19.9$). As RR~Lyrae stars have 
reasonably
well-defined luminosity, that author or any subsequent reader could have
remarked that AR~CVn was almost certainly the most distant known isolated
star, and, at
least at that time, far beyond the normally acknowledged boundaries of the
Galaxy. Yet to our knowledge there has never after its discovery been another
discussion or report in the literature of any observation of this object.

As the published light curve (Meinunger 1972) for AR CVn is
of only modest quality, and there is no published spectrum,
the conclusion that this is actually an
RR~Lyrae star could be questioned. We have obtained photometric and
spectroscopic observations of the variable, which prove to confirm the
remarkable properties of this star.
There are a handful of halo stars known or suspected to lie at comparable or 
greater distances, although we believe this to be the most distant RR~Lyrae 
that is fully characterized with spectrum, period, and light curve.

Accurate coordinates are given in Table~1.
In Figure~3 we show the spectrum of AR~CVn, obtained on UT
1998 January 30. The spectrum was by poor fortune obtained near the faintest
state of the variable, but despite the limited signal, it is clear that this 
is
an F-type star with a weak Ca~K line, entirely consistent with the RR~Lyrae
classification originally proposed.
 
We have obtained photometry of AR~CVn on 10 nights during 1998
January - April, largely from quite brief (100~sec) CCD images made during
unrelated programs. On a few nights multiple observations separated by
hours were available. DIS provides simultaneous images in two broadband
colors, but the red data on this relatively faint object are of modest
quality, and thus that light curve is
not discussed here.
We find that the
object varies over a range of 1.15~mag in the blue,
in reasonable agreement with the
photographically-determined results (Meinunger 1972).
Marked variations on timescales of one hour are
quite evident. Phase dispersion minimization analysis reveals a strong
periodicity of $0.62717\pm 0.00003$~day, again in good agreement with the
discovery result. The light curve folded on this period is shown in Figure~4,
and is of common shape and amplitude.
Phase zero (maximum light) occurs at HJD~2450892.999.
In all respects other than its faint magnitude, AR~CVn appears from our
spectroscopy and photometry to be a perfectly normal RR~Lyrae star, of Bailey
type~{\it ab}.
 
Both the blue and red data are in a broad wavelength band
peculiar to this instrument, but have been transformed to the traditional
Johnson system via observation of AR~CVn as well as spectrophotometric
standard stars in more standard filters on a night of photometric quality.
Although this transformation doubtless introduces an additional uncertainty of
$\sim0.1$~mag beyond that of photon statistics,
issues such as the period, amplitude, and shape of
the light curve of the variable are obviously not affected.
We derive $V_{min} = 19.28$ for the best calibrated data. As RR~Lyrae
stars typically
display $(B-V)_{min}=0.5$, our data agree well with the
$B_{min}=19.9$ quoted by Meinunger. As those data have the virtue of lying in
a standardized bandpass and are supported by our CCD
observations, we adopt his value $<B>=19.4$ for the remainder of
the discussion.
 
In the past decade a variety of surveys have revealed modest numbers of faint
RR~Lyrae stars (Saha 1984, Wetterer et al. 1996), and several dozen objects as
faint as $<B>\sim19$ are now cataloged. However, we are aware of less than a
handful of RR Lyrae stars as faint or slightly fainter than AR~CVn (Hawkins 
1983, 1984; Ciardullo et al. 1989),
and indeed only one of these has a period
determined; none have both period and spectrum as we now present for AR~CVn.
Thus we believe this
interesting object needs to be brought forward again for very practical
current problems, as well as its historical import.
 
Inferences on the luminosity of RR~Lyrae stars from HIPPARCOS data 
have recently been
summarized by Fernley et al. (1998a). There is a weak and uncertain
dependence on the metallicity of the star ({\it e.g.}, Fernley et al. 1998b
and references therein), which we cannot determine with accuracy from our
spectrum for this relatively faint object, other than to note that the
strength of the observed Ca~K would hint that this object is not extremely 
metal poor, perhaps somewhat surprising given its position in the Galaxy.
While our spectroscopic data are of insufficient quality for a meaningful
determination of the metallicity-related $\Delta S$ parameter, this would
clearly be of interest. We therefore arbitrarily 
adopt [Fe/H] = $-1.2$, corresponding to
the metal-rich extreme of halo globular clusters, implying $M_V=0.83$ if the
Fernley et al. (1998a) luminosities are adopted. We
assume negligible interstellar extinction at this very high galactic latitude,
in agreement with the conclusion by numerous investigators that the nearby
cluster M3 has negligible reddening. We thus infer a heliocentric distance of
40~kpc for AR~CVn, and the corresponding galactocentric distance for this
position near the galactic pole is very similar.
The RR~Lyrae luminosity calibration of Gould \& Popowski (1998 and references
therein) yields a slightly larger distance, corresponding to an increased
modulus of $0.1-0.2$~mag.
A total uncertainty of 20\%
in our estimate probably encompasses the uncertainties in both the distance
scale and our photometry. Great care should be taken when comparing this
distance estimate with those published for other very faint RR~Lyrae stars
over the years, as estimates of their luminosity and its metal dependence
have fluctuated substantially over time; direct comparison of mean apparent
magnitudes of the stars is of course free of this complication.
 
Our spectroscopic observations were not optimized for absolute velocity
determination, but using the average of the centroids of four Balmer lines 
plus Ca~H\&K, we measure heliocentric velocity $v=55\pm 60$ km~s$^{-1}$.
For the observed photometric variability amplitude, the correction to the
radial velocity for pulsation (Ledoux \& Walraven 1958) is considerably 
smaller than
our measurement errors, and is therefore neglected. The correction to the LSR
(+11~km~s$^{-1}$) and the radial component of the LSR in this direction
($\sim+30$~km~s$^{-1}$) are also both small, and for this geometry near the
galactic pole, the radial velocity of AR CVn with respect to the Sun/LSR
is similar to its galactocentric radial velocity. Therefore it is plausible
that AR~CVn is bound to the Galaxy.

The one comparably distant halo RR~Lyrae of which we are aware where a
velocity is published, R15, was initially attributed with an extreme high
velocity ($-465$~km~s$^{-1}$) by Hawkins (1983), but then revised to a much
lower value  ($-218$~km~s$^{-1}$) by Norris \& Hawkins (1991). As the inferred
enclosed mass scales as the square of the velocity of a probe,
these disparate velocities stand as a warning
regarding the dangers of small sample size when addressing the kinematics of
the very distant halo.
 
Although one may from relatively local observations infer indirectly the
existence of a very extended stellar halo to our Galaxy (Carney et al. 1988),
direct observation and characterization of distant stars in this halo is
difficult. Aside from the faint RR~Lyrae stars discussed here, at least two
additional stellar tracers of the very distant halo are available, the faint
high latitude carbons stars (Totten \& Irwin 1998), some perhaps as distant
as 100~kpc (Margon et al. 1984), and faint blue horizontal branch stars
(Norris \& Hawkins 1991). However, the RR~Lyrae stars surely have the most
confidently determined distances of the various tracers. Some undetermined
fraction of faint halo C stars, for example, are in fact very nearby dwarfs
(Green \& Margon 1994, Green 1998).

Although in principle it is possible to use even a single distant halo star to
estimate the enclosed mass of the Galaxy (Hawkins 1983), we believe that at
least a modest-sized sample is needed for such estimates to be plausible. Such
a sample of RR~Lyrae's is slowly becoming available, and should be greatly 
accelerated by the {\it Sloan Digital Sky Survey} (Margon 1998), where large 
numbers of faint RR~Lyrae's will be identified (Krisciunas et 
al. 1998).

\acknowledgments

We thank K.~Gloria, T.~Hoyes, G.~Magnier, B.~Skelton,
C.~Stubbs, and E.~Turner
for their aid in obtaining the observations of AR~CVn, and G.~Wallerstein for
useful discussions on that object. S.~Anderson, B.~Beck-Winchatz, and
A.~Diercks kindly provided images of S~10765. This research has made use of
the SIMBAD database, operated at CDS, Strasbourg, France.


\clearpage
\begin{deluxetable}{llclrc}
\tablenum{1}
\tablecolumns{6}
\tablecaption{Astrometric, Radio, and X-ray Data for ``Nicht Rote" Objects}
\tablehead{
\colhead{Meinunger} &
\colhead{} &
\colhead{} &
\colhead{} &
\colhead{FIRST\,\tablenotemark{a}} &
\colhead{RASS\,\tablenotemark{b}} \nl
\colhead{Name} &
\colhead{Other Name} &
\colhead{$\alpha$ (J2000) $\delta$} &
\colhead{$z$} &
\colhead{(mJy)} &
\colhead{(cps)}
}
\startdata
S 10760 & AR CVn                    & 13 42 07.16 \ +29 45 55.5 &\ \ \ -- & $<0.93$ &  --  \nl
S 10762 & CC Boo\,\tablenotemark{c} & 13 40 22.84 \ +27 40 58.5 & 0.172 & $<0.90$ & 0.16 \nl
S 10763 & CD Boo                    & 13 41 23.34 \ +27 49 55.4 & 1.045 &   1.21  &  --  \nl
S 10764 & B2 1339+28                & 13 42 10.90 \ +28 28 46.7 & 0.33\tablenotemark{d}  &   1.58  &  --\,\tablenotemark{e}  \nl
S 10765 & NGP9 F324-0276706         & 13 46 47.38 \ +29 54 21.0 & 0.063 & $<0.97$ &  --  \nl
\enddata
\tablenotetext{a}{\,20 cm radio data from White et al. (1997)}
\tablenotetext{b}{\,soft X-ray data from ROSAT All Sky Survey Bright Source Catalog
  (Voges et al. 1998)}
\tablenotetext{c}{\,observations from Margon \& Deutsch (1997)}
\tablenotetext{d}{\,Harris et al. (1992)}
\tablenotetext{e}{\,known soft X-ray emitter (see text)}
\end{deluxetable}

\clearpage

\begin{figure}
\plotone{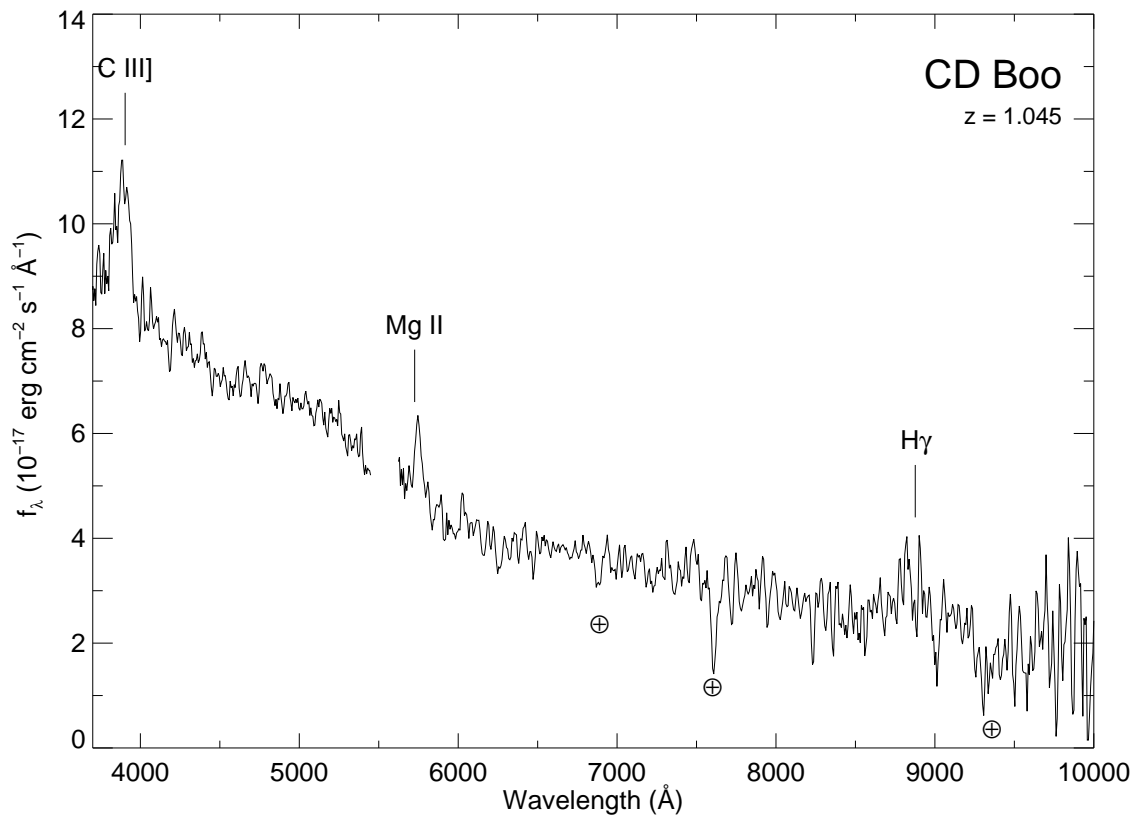}
\caption{
The spectrum of CD~Boo = S~10763, totaling 1680~sec of integration. 
The data for this 
and subsequent spectra in this paper have been 
reduced approximately to absolute fluxes via observations of 
spectrophotometric standard stars, but uncertain light losses at the slit 
imply a $\pm0.3^m$ uncertainty in this calibration. Note the marked UV excess,
and emission at C\,III], Mg\,II, and $H\gamma$. The odd $H\gamma$ profile most
probably results from an intrinsically broad line mutilated by multiple
telluric features. The missing 100\,\AA\ area near 5500\,\AA\ is due to data
marred by the dichroic optic in the spectrograph. This spectrum, as well as
that in Figure~2, has had a 3 pixel smoothing function applied.
}
\end{figure}

\begin{figure}
\plotone{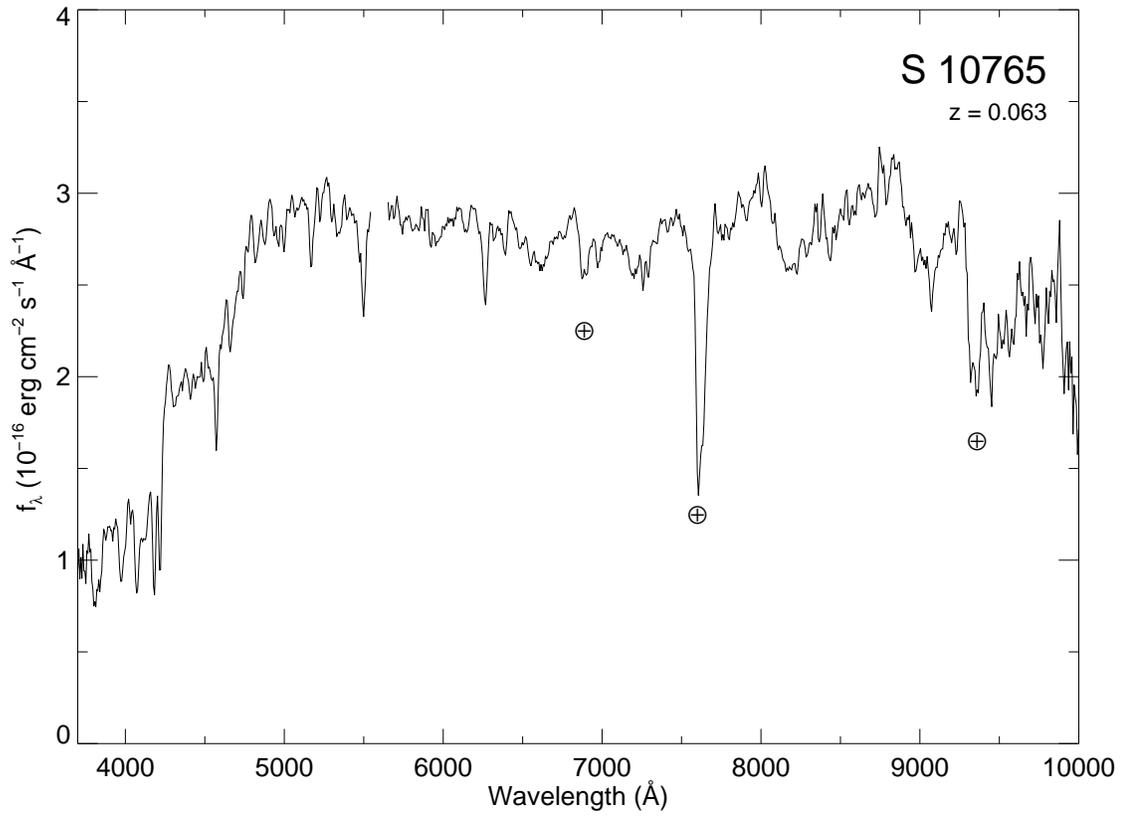}
\caption{
The spectrum of the highly variable galaxy S~10765, derived from 1200~sec of
integration. Note the absence of 
emission lines or a diluting nonthermal continuum, despite the strong evidence
for an active nucleus implied by the variability. 
}
\end{figure}

\begin{figure}
\plotone{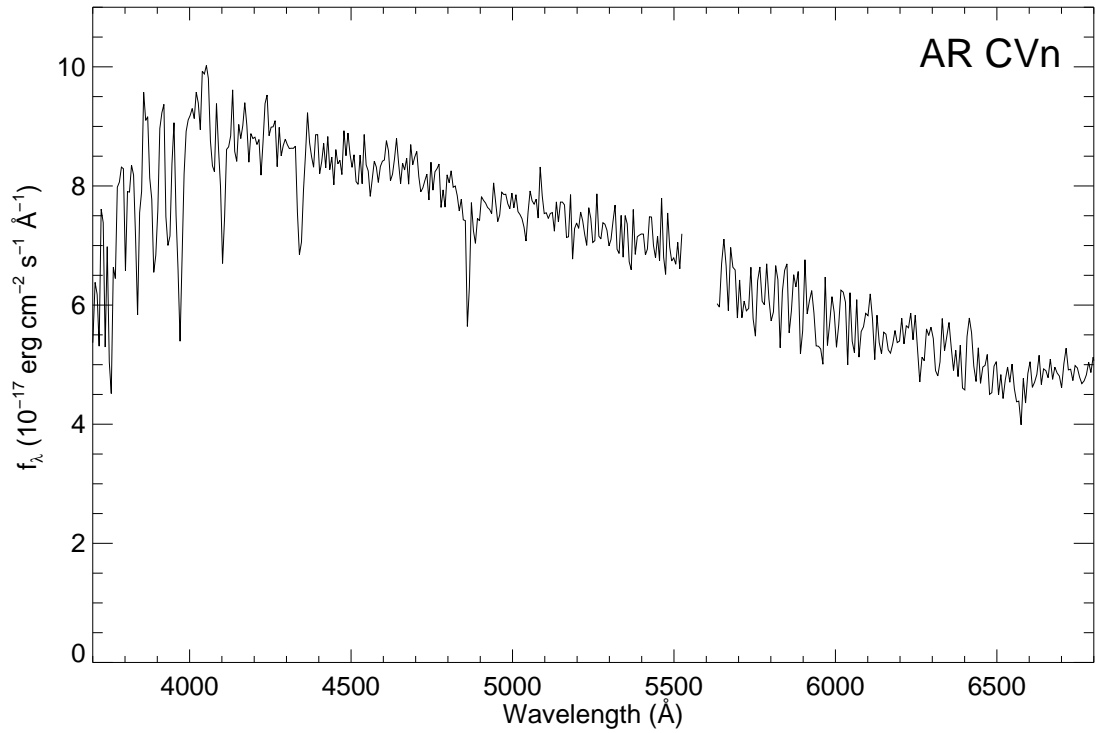}
\caption{
The spectrum of AR CVn, obtained near the phase of minimum light of the
variable. Two exposures of 1,200~sec each were coadded to obtain these data.
The prominent Balmer absorption lines of the F-type spectrum are evident.
}
\end{figure}

\begin{figure}
\plotone{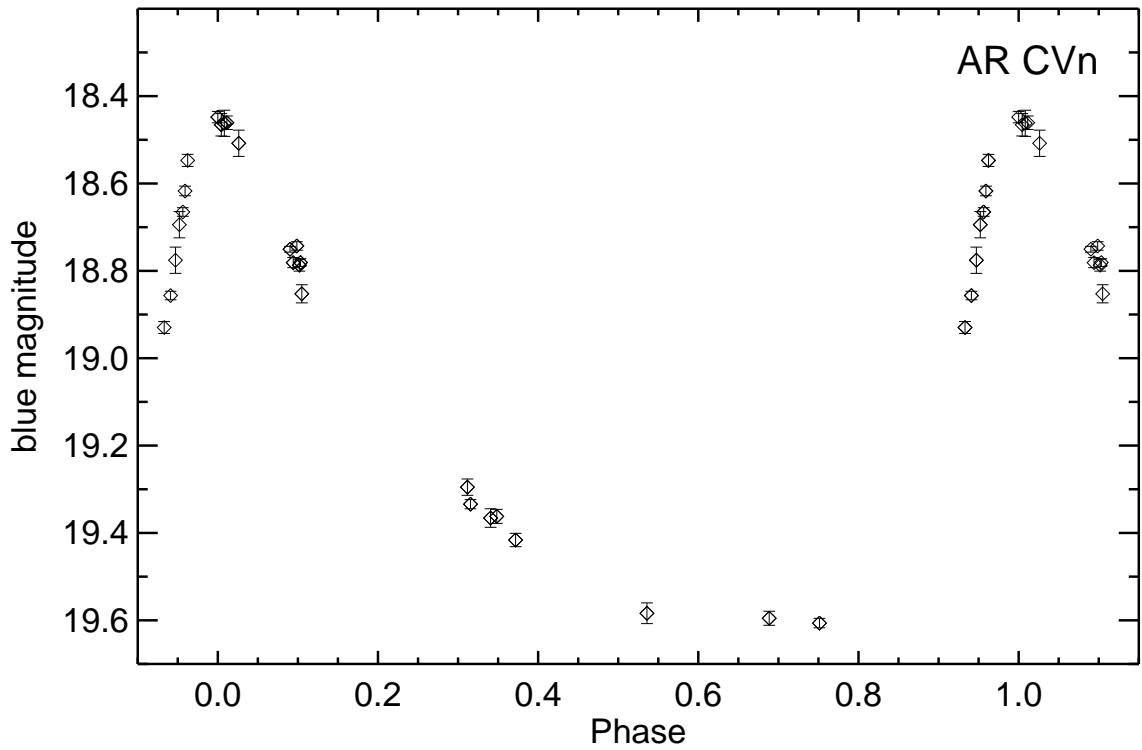}
\caption{
The light curve of AR CVn, folded with the best fit period of
0.62717~day, obtained from observations on 10 nights spanning more than 2
months. Slightly more than one cycle is plotted for clarity. The error bars
reflect relative count-rate uncertainties, but there is no allowance for
systematic offsets of the photometric bandpass, which is non-standard, from
the more common systems such as the Johnson B band. As the shape, period, and
amplitude of the variation clearly confirm this object as an RR~Lyrae, this
further uncertainty is not of great concern.
}
\end{figure}

\clearpage

\end{document}